\documentclass[aps,pra,floatfix,twocolumn]{revtex4}
\usepackage{amsfonts, amsmath, amssymb}
\usepackage{bbding}
\usepackage{amssymb}
\usepackage{color}
\usepackage{amsmath}
\usepackage{graphicx}
\usepackage{subfigure}
\usepackage{txfonts}
\usepackage{bm}
\usepackage{ulem}
\usepackage{slashed}
\usepackage[colorlinks,linkcolor=blue,citecolor=blue]{hyperref}
\usepackage[utf8]{inputenc}
\newcommand{\nn}[1]{{\text{NN}}}
\newcommand{\nnn}[1]{{\text{NNN}}}

\begin{document}

\title{Steady-state phases of dissipative spin-$1/2$ XYZ model with frustrated interaction}
\author{Xingli Li, Yan Li, and Jiasen Jin}
\email{jsjin@dlut.edu.cn}
\affiliation{School of Physics, Dalian University of Technology, 116024 Dalian, China}
\date{\today}
\begin{abstract}
We investigate the steady-state phases of the dissipative spin-1/2 $J_{1}$-$J_{2}$ XYZ model on a two-dimensional square lattice. We show the next-nearest-neighboring interaction plays a crucial role in determining the steady-state properties. By means of the Gutzwiller mean-field factorization, we find the emergence of antiferromagnetic steady-state phases. The existence of such antiferromagnetic steady-state phases in thermodynamic limit is confirmed by the cluster mean-field analysis. Moreover, we find the evidence of the limit cycle phase through the largest quantum Lyapunov exponent in small cluster, and check the stability of the oscillation by calculating the averaged oscillation amplitude up to $4\times4$ cluster mean-field approximation.
\end{abstract}
\maketitle
\section{Introduction}
\label{Introduction}
Quantum phase transition under equilibrium conditions has achieved a profound understanding in the past decades. The quantum phase transition is manifested by the continuous or abrupt changes of the ground state of a quantum many-body system when varying the external parameter. The spontaneously symmetry broken in the ground state is essentially driven by the quantum fluctuations \cite{SSachdev}.

The phase transition in quantum many-body system may also happen under the out-of-equilibrium condition. Actually, the inevitably interactions of a quantum system and its environment always drive the system, referred to as open system, far from equilibrium. Because the thermal equilibrium is absent, the stationary property of the nonequilibrium system is determined by the asymptotical steady state of the nonunitary dynamics in the long-time limit. Usually, the dynamics of the open system in a Markovian (memoryless) environment is well described by the quantum master equation in Lindblad form $\dot{\hat{\rho}}(t)=\mathcal{L}\hat{\rho}(t)$ where $\mathcal{L}$ is the so-called Liouvillian superoperator \cite{Lindblad1976, Gorini1976,breuer_book}. The properties of the steady state, as a result of the competition between the coherent evolution and the dissipative process, can be captured by the spectrum of $\mathcal{L}$ \cite{Minganti2018}. Analogous to the equilibrium case, the steady-state symmetry ruled by the Lindblad master equation may also be spontaneously broken in the thermodynamic limit.

The steady-state phase diagram of the open quantum many-body system is predicted to be particularly rich \cite{TonyELee2013PRL,VSPRA2017,JiasenJinPRB2018,DolfHuybrechtsPRB2020}. It displays exotic phases that spontaneously broken the symmetries possessed by the Liouvillian of the system \cite{TonyELee2011PRA,TonyELee2013PRL,JiaSenJin2013PRL,Hendrik2015PRL,Vincent2017PRB,Landa2020PRB}. Among the steady-state phases, the limit cycle (LC) phase which spontaneously breaks the time translation invariance has attracted significant attention  \cite{LudwigPRL2013,WenlinPRE2016,MinghuXu2014PRL,TonyELee2014PRE,DTNJP2018}. It is considered to be a potential realization of time crystals in nonequilibrium system \cite{ZGong2018PRL,FIemini2018PRL,KristopherTucker2018NJP}. Experimental investigations of nonequilibrium properties of open quantum many-body systems have been realized in trapped ions \cite{Zoller2012}, ultracold atomic gases in optical lattices \cite{Greiner2002,Bloch2008PRL,KBaumann2011PRL,KBaumann2010Nature,bluvstein2021}, and arrays of coupled QED cavities \cite{AndrewAHouck2012NP,MattiasFitzpatrick2017PRX,collodo2019}.

Recently, it is explored that the frustration in many-body system can induce fantastic nontrivial steady-state properties, such as the antiferromagnetism, spin-density wave and chaotic dynamics \cite{JingQian2013PRA,XingLi2021PRB,qiao2020,ZejianPRA2021}. The frustration refers to the fact that the competing interactions between neighboring sites cannot be satisfied simultaneously \cite{FigueiridoPRB1990,DaisukeYamamoto2014PRL,SimengYan2011Sci}. Generally, the presence of frustration is characterized by a large degeneracy in ground-state energy \cite{WannierPR1950}. It is believed that the frustration tends to destroy conventional long-range orders. Basically, the frustration stems from either the geometry of the lattice or the competition among interactions in the system. We call the former the geometrically frustration while the latter interaction frustration. One of the prototypes of the geometrically frustrated system is the two-dimensional Ising antiferromagnet on a triangular lattice. In this well-known model, the incompatible antiferromagnetic interplay emerges once two of the spins are aligned oppositely to satisfy the antiferromagnetic interaction and the third one can not be antialigned to the other two spins simultaneously. The macroscopically degenerated ground state shows the fluid-like behavior \cite{WenNPJQ2019,zhou2017}. The geometrical frustration have been realized experimentally \cite{KKim2010Nature,AndreEckardt2010EPL,CBecker2010NJP}. Theoretically, it has been shown that in the geometrically frustrated spin-1/2 system on a triangular lattice unconventional steady-state antiferromagnetism and spin-density wave emerge \cite{JingQian2013PRA,XingLi2021PRB}.

Regarding to the competing-interaction frustrated system, the typical example is the $J_1$-$J_2$ spin-1/2 Heisenberg model on square lattice in which both the nearest- ($J_1$) and next-nearest- ($J_2$) neighboring interactions are considered. The competition between the nearest-neighboring (NN) and next-nearest-neighboring (NNN) interactions dramatically modifies the Hamiltonian spectrum of the system with only NN interactions. The ratio $J_1/J_2$ determines the properties of the ground state of the system. In particular, for $J_1/J_2<1$, the ground state in the so-called N\'eel state, while in the opposite side $J_1/J_2>1$, the ground state is in the collinear striped antiferromagnetic (CAF) order \cite{OGPRB2012,ZengPRB2010}. In the intermediate region, the long-range order is suppressed by the quantum fluctuations and the system is strongly frustrated. It is believed that the quantum spin liquid may exist in this region \cite{WenyuanXiv2020}.

Inspired by the rich ground-state phase diagram induced by the competing NN and NNN interactions in the equilibrium case, we are going to investigate the steady-state properties of the open quantum many-body system with the $J_1$-$J_2$ interactions.
As a concrete example, we focus on a dissipative spin-1/2 XYZ model on two-dimensional square lattice. We consider both the NN and NNN anisotropic Heisenberg couplings among the sites. In addition, the local dissipative processes on each site that drive the system out-of-equilibrium are considered taken into account. Our goal is to discover the novel steady-state phases that is brought by the NNN couplings. By employing a combination of the state-of-the-art approaches, we shed light on the impact of NNN interactions in determining the steady state. As the NNN coupling varying, the system exhibits various steady-state phases. We mostly concentrate on the antiferromagnetic phases. We also predict a LC phase in which the steady state is time periodic. In particular, the emergence of the LC phase is highlighted by the largest quantum Lyapunov exponent (LE) and averaged oscillation amplitude. The existence of LC phase is closely connected to the dissipative time crystals \cite{FIemini2018PRL,KristopherTucker2018NJP,fan2020,seibold2020,landa2020,prazeres2021}.

\begin{figure*}[!htp]
\includegraphics[width=0.9\textwidth]{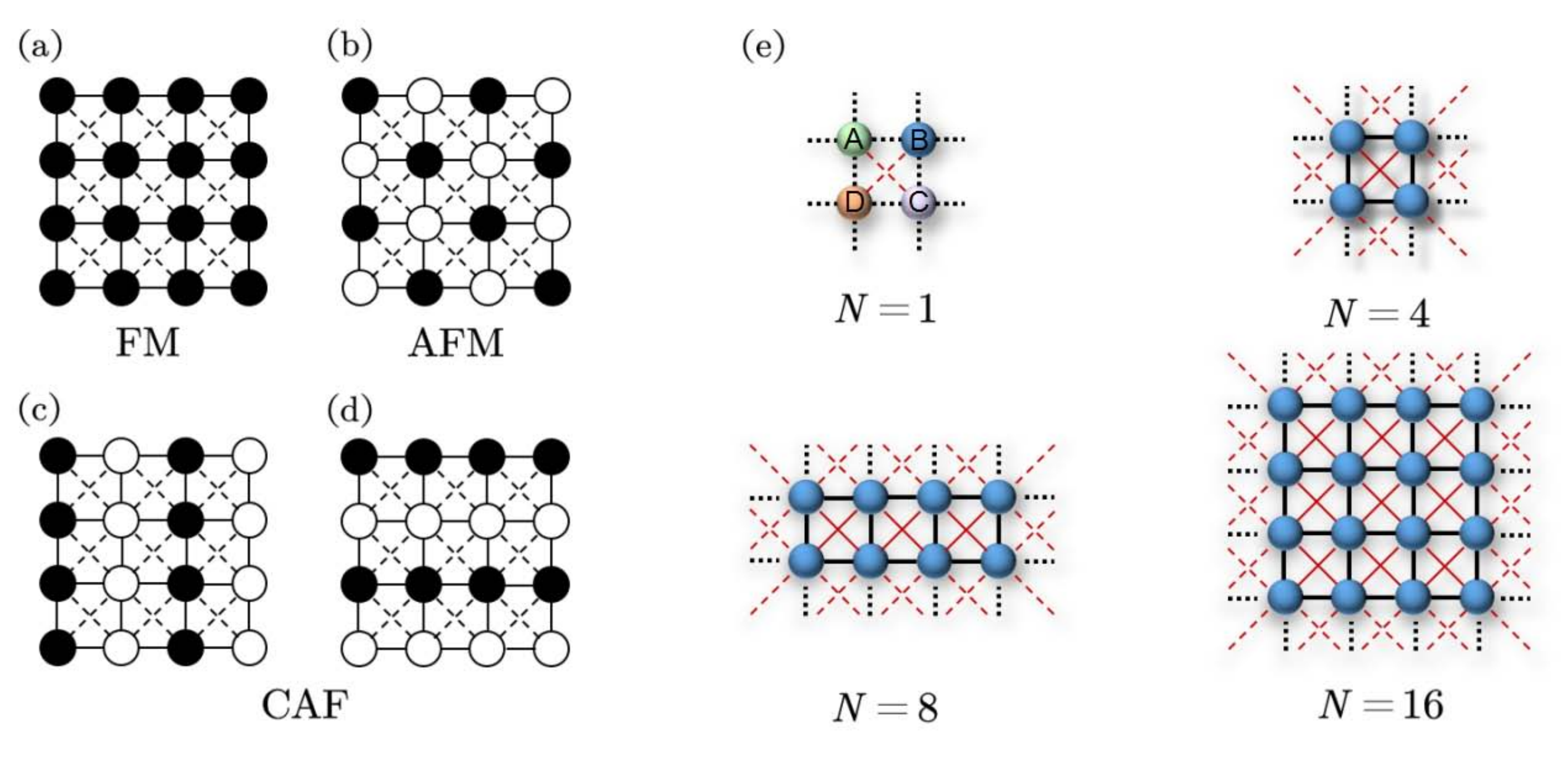}
\caption{\label{fig1} (Color online) (a)-(d) Illustration of partial $\mathbb{Z}_{2}$-symmetry broken quantum phases in Heisenberg XYZ model, the black and white colors represent the different magnetization directions in $x$-axis. (a) ferromagnetic (FM) order; (b) antiferromagnetic (AFM) order; (c)-(d) collinear striped antiferromagnetic (CAF) order. (e) Schematic diagram of the Gutzwiller factorization of the full lattice. In the mean-field approximation $(N=1)$, the interactions between sites are all treated as an effective field (the dashed bonds), the lattice are divide into four sublattices (in different colors), marked by $A$, $B$, $C$ and $D$. For the cluster mean-field approximation with different sizes $(N \geq 4)$, the interactions between the sites inside the cluster (the solid bonds) are treated exactly, the NN and NNN interactions outside of the cluster are treated as an effective field.}
\label{illustration}
\end{figure*}

This paper is organized as follows. In Sec.\ref{Sec:Model}, we explain the dissipative spin-1/2 $J_1$-$J_2$ XYZ model on the square lattice and the corresponding master equation that describes the evolution of the system. We present the possible steady-state phases that may appear in the system. In Sec.\ref{Sec:Mean-field approximation}, by employing the Gutzwiller single-site mean-field (MF) factorization, we solve the steady-state solutions to the single-site MF master equation. By performing the linear stability analysis on the MF fixed points, we uncover the various steady-state phase. In Sec.\ref{Sec:Cluster Mean-field Method}, we include the short-range interaction in the discussion by performing the cluster mean-field (CMF) method. We confirm the existence of the CAF phase and show the evidence of the LC phase through the largest LE and the average oscillation amplitude. We summarize in Sec.\ref{Sec:Conclusions}.

\section{Model}
\label{Sec:Model}

The model we consider here is a spin-1/2 quantum many-body model on square lattice whose Hamiltonian is given by (set $\hbar=1$ hereinafter),
 \begin{equation}
\hat{H} = \sum_{\alpha}{J_\alpha\left[J_{1}\sum_{\langle j,l\rangle}\hat{\sigma}^{\alpha}_{j}\hat{\sigma}^{\alpha}_{l}+J_{2}\sum_{\langle\langle j,l\rangle\rangle}\hat{\sigma}^{\alpha}_{j}\hat{\sigma}^{\alpha}_{l}\right]},
\label{Total Hamiltonian}
\end{equation}
where $\hat{\sigma}^{\alpha}_{j} (\alpha=x,y,z)$ are the Pauli matrices for the $j$-th site, $\langle j,l\rangle$ and $\langle\langle j,l\rangle\rangle$ denote the sums of the spin-spin coupling run over the nearest-neighboring and next-nearest-neighboring interactions, respectively. $J_\alpha$ are the coupling constants. For $J_{2}=0$,  we recover the conventional XYZ model with NN couplings. The XYZ Hamiltonian is generic in spin systems and can be reduced to the isotropic Heisenberg Hamiltonian for $J_x=J_y=J_z$ and Ising Hamiltonian for $J_x=J_y=0$.

In order to drive the system out-of-equilibrium, we assume that each spin contacts with a Markovian environment which leads to a local dissipative process on each spin. In our specific model, the local environment tends to incoherently flip each spin down to the $z$-direction. Thus the quantum master equation governing the evolution of the system's density matrix $\hat{\rho}(t)$ is
\begin{equation}
\frac{d\hat{\rho}}{dt}(t)=\mathcal{L}[\hat{\rho}(t)]=-i[\hat{H},\hat{\rho}(t)] + \sum_{j}\mathcal{D}_{j}[\hat{\rho}(t)],
\label{Lindblad Master Equation}
\end{equation}
where $\mathcal{L}$ is the Liouvillian superoperator. The local dissipator $\mathcal{D}_j$ on the $j$-th site takes the form of
\begin{equation}
\mathcal{D}_{j}[\hat{\rho}(t)] = \frac{\gamma}{2}\left[2\hat{\sigma}^{-}_{j}\hat{\rho}(t)\hat{\sigma}^{+}_{j} - \hat{\sigma}^{+}_{j}\hat{\sigma}^{-}_{j}\hat{\rho}(t)-\hat{\rho}(t)\hat{\sigma}^{+}_{j}\hat{\sigma}^{-}_{j}\right],
\label{Lindbladian}
\end{equation}
where $\gamma$ is the decay rate and the operators $\hat{\sigma}^{\pm}_{j} = (\hat{\sigma}^{x}_{j} \pm i\hat{\sigma}^{y}_{j})/2$ represent the raising and lowering operators for the $j$-th spin. In following, we will always work in units of $\gamma$. Additionally, for simplicity, we set $J_{1}/\gamma=1$ and restrict the NNN coupling to be $J_{2}/\gamma\in (0,1)$ in this work.

The Lindblad master equation Eq. (\ref{Lindblad Master Equation}) admits the $\mathbb{Z}_{2}$ symmetry associated to a $\pi$ rotation of all the spins about the $z$-axis $(\hat{\sigma}^{x}_{j}\to-\hat{\sigma}^{x}_{j},\hat{\sigma}^{y}_{j}\rightarrow-\hat{\sigma}^{y}_{j}, \forall j)$. In the thermodynamic limit, this $\mathbb{Z}_{2}$ symmetry may be spontaneously broken as the strengths of spin-spin interactions varying. In the symmetry-broken phases the magnetization on the $x$-$y$ plane of each spin is nonzero and could be spatially modulated. Here we list the possible steady-state phases as the following,

(i) {\it Paramagnetic (PM) phase}. This is a trivial uniform state in which all the spins are pointing down along $z$-axis, $\langle\hat{\sigma}^{x}\rangle = \langle\hat{\sigma}^{y}\rangle = 0$ indicating that the system preserves the $\mathbb{Z}_{2}$-symmetry. The notation $\langle\hat{\sigma^\alpha\rangle} = \text{tr}(\sigma^\alpha\rho)$ ($\alpha=x,y,z$) means the expectation value of $\hat{\sigma}^\alpha$.

(ii) {\it Ferromagnetic (FM) phase}. The FM phase is a uniform ordered phase. Each spin has an identical nonzero steady-state magnetization on the $xy$ plane, namely $\langle\hat{\sigma}^{x}\rangle\neq 0,\langle\hat{\sigma}^{y}\rangle \neq 0$ as shown in Fig.\ref{fig1}(a), indicating that the $\mathbb{Z}_{2}$-symmetry is broken.

(iii) {\it Antiferromagnetic (AFM) phase}. The AFM phase is a nonuniform ordered phase. In the AFM phase, as shown in Fig.\ref{fig1}(b), the whole lattice is divided into two alternating sublattices. All the spins have a nonzero steady-state magnetizations on the $x$-$y$ plane. Moreover, the spin on one sublattice points to a different direction to the other. The steady-state magnetization is spatially modulated with a period of twice the lattice constant.

(iv) {\it Collinear striped antiferromagnetic (CAF) phase}. The CAF phase is another type of nonuniform ordered phase. In the CAF phase, the spins on the lattice are collinearly polarized. The steady-state magnetization is spatially modulated with a period of twice lattice constant in either $x$ or $y$ direction as shown in Fig.\ref{fig1}(c)-(d).

\subsection{The frustration in the $J_1$-$J_2$ XYZ Hamiltonian}
In this subsection, we check the existence of frustration in the studied Hamiltonian Eq. (\ref{Total Hamiltonian}). We adopt the measure of frustration proposed in Refs. \cite{giampaolo2011,marzolino2013}
which quantifies the incompatibility between the global and local orders. A many-body Hamiltonian can be expressed as $\hat{H}_G = \sum_{\ell}{\hat{h}_{\ell}}$ where $G$ stands for the global system and $\ell=\langle i,j\rangle$ or $\langle\langle i,j\rangle\rangle$ stands for the subsystem associated with local interactions $\hat{h}_{\ell}$. The measure of frustration for $\hat{h}_\ell$ is defined as follows
\begin{equation}
f_\ell = 1 - \text{tr}\left[\hat{\rho}_\ell\hat{\Pi}_\ell\right],
\label{fl}
\end{equation}
where $\hat{\rho}_\ell=\text{tr}_{\ne\ell}\hat{\rho}_G$ is the reduced local state obtained from the partial trace of the global ground state $\hat{\rho}_G$ of $\hat{H}_G$ over the rest of the system, $\hat{\Pi}_\ell$ is the projector onto the ground-state space of the local Hamiltonian $\hat{h}_\ell$. The second term on the r.h.s. of Eq. (\ref{fl}) quantifies the overlap between the reduced local state and the local ground state associated to $\hat{h}_\ell$. Therefore the system is frustration-free if $f_\ell=0$, $\forall \ell$. The total frustration of the global Hamiltonian is thus defined by averaging over all the local measures $f_\ell$. This measure quantifies the frustrations due to the geometry of the system, the competing interactions, and the noncommutativity between different $\hat{h}_\ell$s.

\begin{figure}[!htp]
\includegraphics[width=0.45\textwidth]{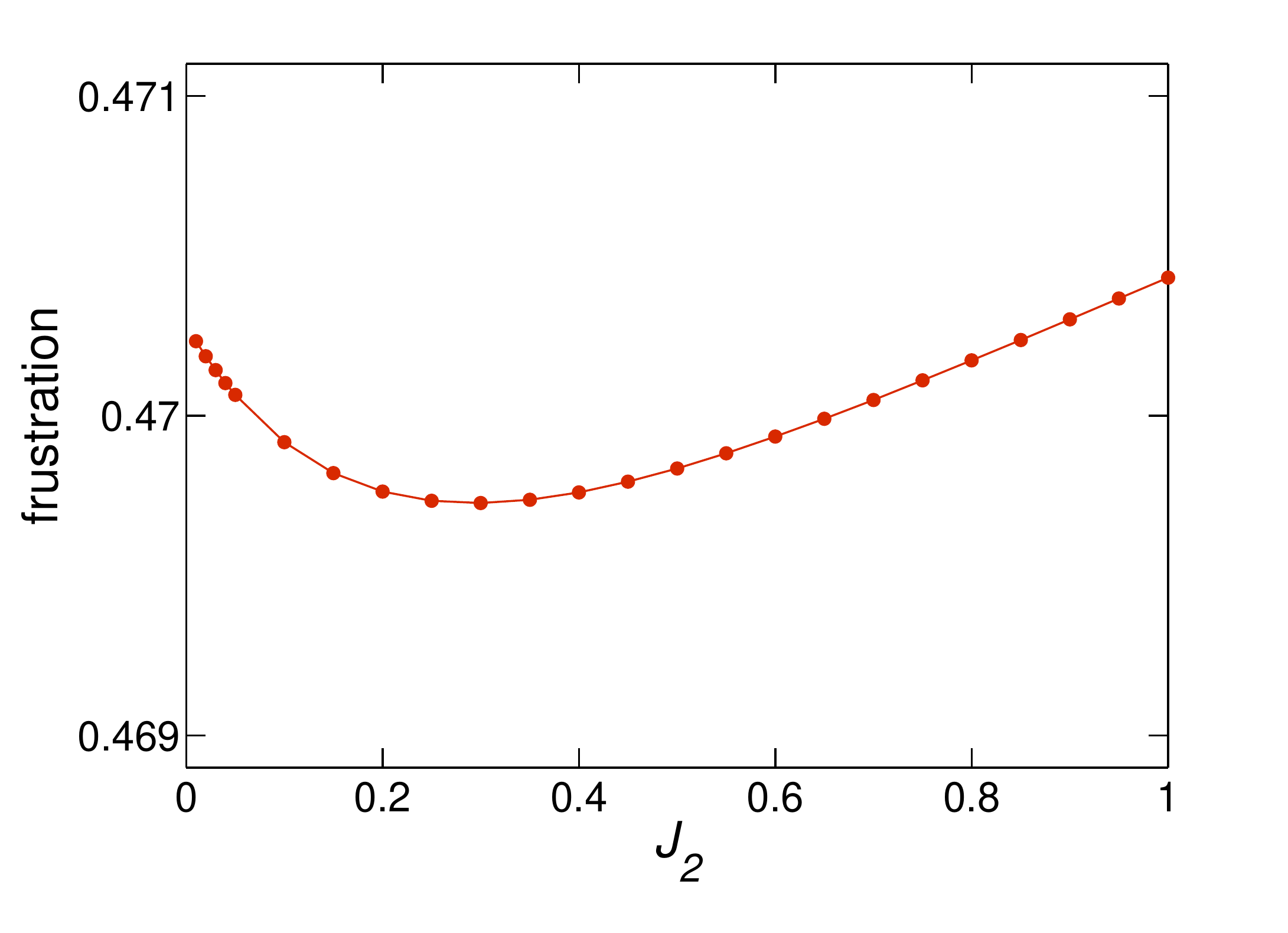}
\caption{\label{fig5}(Color online) The total frustration as a function of $J_{2}$ for the XYZ Hamiltonian on a $4\times4$ cluster (open boundary condition). The parameters are chosen as $\{J_x,J_y,J_z\} = \{-3.2,-1,1\}$. }
\label{frus}
\end{figure}

The effects of frustrations in the XYZ model with competing $J_1$-$J_2$ interactions have been discussed in Ref. \cite{giampaolo2015}. Here, in Fig. \ref{frus}, we show the total frustration as a function of the strength of the NNN coupling of the XYZ model on $4\times4$ lattice (open boundary condition). For the chosen parameters, one can see that the frustration is always present although the strength of NNN coupling affects the quantity of the frustration slightly. As will be seen in Sec. \ref{Sec:Cluster Mean-field Method} , the system exhibits various steady-state phases. Note that because the global ground state is two-fold degenerate for the specific parameters, $\rho_G$ is taken as the equiprobable statistical average of the two degenerate global ground states, namely the maximally mixed global ground state \cite{marzolino2013}.

\section{Mean-field approximation}
\label{Sec:Mean-field approximation}
Due to the complexity of the full quantum master equation, we start with the single-site MF method basing on the Gutzwiller factorization. The density matrix for the whole lattice is factorized as $\hat{\rho}=\bigotimes_{j}\hat{\rho}_{j}$ with the reduced density matrix $\rho_j=\text{tr}_{\ne j} \rho$ for each site. The reduced density matrices belong to the same sublattice are assumed to be identical. Substituting the factorized density matrix into Eq. (\ref{Lindblad Master Equation}), we may obtain the single-site MF master equation for each sublattice in the following form,
\begin{equation}
\frac{d\hat{\rho}_{j}}{dt}=-i[\hat{H}^{\text{mf}}_j,\hat{\rho}_{j}] + \frac{\gamma}{2}\left[2\hat{\sigma}_j^{-}\hat{\rho}_{j}\hat{\sigma}_j^{+}-\{\hat{\sigma}_j^{+}\hat{\sigma}_j^{-},\hat{\rho}_{j}\}\right],
\label{eq:mfmasterequation}
\end{equation}
where $j = A, B, C$ and $D$ denotes the sublattice. The corresponding MF Hamiltonian for sublattice $j$ is governed by
\begin{equation}
\hat{H}^{\text{mf}}_j = \sum_{\alpha = x,y,z}{\sum_{k,l} J_{\alpha}\hat{\sigma}_j^{\alpha}(J_{1}\langle\hat{\sigma}^{\alpha}_{k}\rangle + J_{2}\langle\hat{\sigma}^{\alpha}_{l}\rangle}),
\label{mf_Hamiltonian}
\end{equation}
where $\langle\hat{\sigma}^\alpha_{j,l}\rangle = \text{tr}(\hat{\sigma^\alpha}\hat{\rho}_{j,l})$, and the subscripts $k$ and $l$ denote the nearest and next-nearest neighbors of site $j$, respectively. By virtue of Eqs. (\ref{eq:mfmasterequation}) and (\ref{mf_Hamiltonian}),  we obtain the following system of Bloch equations for each sublattice as,
\begin{widetext}
\begin{equation}
\begin{aligned}
\frac{d\langle\hat{\sigma}^{x}_{j}\rangle}{dt} =& 2\sum_{l}\sum_{k}J_{y}(J_{1}\langle\hat{\sigma}^{y}_{k}\rangle + J_{2}\langle\hat{\sigma}^{y}_{l}\rangle)\langle\hat{\sigma}^{z}_{j}\rangle - J_{z}(J_{1}\langle\hat{\sigma}^{z}_{k}\rangle + J_{2}\langle\hat{\sigma}^{z}_{l}\rangle)\langle\hat{\sigma}^{y}_{j}\rangle-\frac{\gamma}{2}\langle\hat{\sigma}^{x}_{j}\rangle,\\
\frac{d\langle\hat{\sigma}^{y}_{j}\rangle}{dt} =& 2\sum_{l}\sum_{k}J_{z}(J_{1}\langle\hat{\sigma}^{z}_{k}\rangle + J_{2}\langle\hat{\sigma}^{z}_{l}\rangle)\langle\hat{\sigma}^{x}_{j}\rangle - J_{x}(J_{1}\langle\hat{\sigma}^{x}_{k}\rangle + J_{2}\langle\hat{\sigma}^{x}_{l}\rangle)\langle\hat{\sigma}^{z}_{j}\rangle-\frac{\gamma}{2}\langle\hat{\sigma}^{y}_{j}\rangle,\\
\frac{d\langle\hat{\sigma}^{z}_{j}\rangle}{dt} =& 2\sum_{l}\sum_{k}J_{x}(J_{1}\langle\hat{\sigma}^{x}_{k}\rangle + J_{2}\langle\hat{\sigma}^{x}_{l}\rangle)\langle\hat{\sigma}^{y}_{j}\rangle - J_{y}(J_{1}\langle\hat{\sigma}^{y}_{k}\rangle + J_{2}\langle\hat{\sigma}^{y}_{l}\rangle)\langle\hat{\sigma}^{x}_{j}\rangle-\gamma\left(\langle\hat{\sigma}^{z}_{j}\rangle+1\right),\\
\label{Mean-field equation system}
\end{aligned}
\end{equation}
\end{widetext}
here again the sum over $k$, $l$ are taken over the nearest and next-nearest neighbors of site $j$, respectively. The fixed points can be determined by setting Eq. (\ref{Mean-field equation system}) to be zero. Apparently, the state $\hat{\rho}_{j,\downarrow}=|\downarrow_j\rangle\langle\downarrow_j|$, with the spin pointing down to the $z$-direction, is always a steady-state solution to Eq. (\ref{Mean-field equation system}). The joint state of the whole lattice is thus given by $\hat{\rho}_{\downarrow}=\bigotimes_{j}\hat{\rho}_{j,\downarrow}$ indicating that the system is in the PM phase. However, $\hat{\rho}_{\downarrow}$ is not always stable; the linear stability analysis on $\hat{\rho}_{\downarrow}$ can reveal the possibility of transitions from the PM to other phases.

The idea of linear stability analysis is to introduce local small fluctuations $\delta \rho_{j}$ to around the MF steady state by
\begin{equation}
\hat{\rho}_\downarrow\rightarrow\bigotimes_{j}\left(\hat{\rho}_{j,\downarrow} + \delta \rho_{j}\right),
\end{equation}
and check how the perturbations evolve with time. We expand the perturbations in terms of plane waves
\begin{equation}
\delta \rho_{j} = \sum_{\textbf{k}}e^{-i\textbf{k}\cdot\textbf{r}_{j}}\delta \rho^{\textbf{k}}_{j},
\end{equation}
where $\textbf{k}$ is the wave vector. Thus the equation of motion for the perturbation $\delta\rho^{\textbf{k}}_{j}$ are decoupled in the momentum space and reads,
\begin{equation}
\partial_{t}\delta\rho^{\textbf{k}}=\mathcal{L}_{\textbf{k}}\cdot\delta\rho^{\textbf{k}}.
\end{equation}
The superoperator $\mathcal{L}_{\textbf{k}}$ has the following form,
\begin{equation}
\mathcal{L}_{\textbf{k}} =
\begin{pmatrix}
     -\gamma & 0                   & 0                    & 0 \\
      0      &P -\frac{\gamma}{2} & Q                   & 0 \\
      0      & -Q                 & -P -\frac{\gamma}{2} & 0 \\
      \gamma & 0                   & 0                    & 0
\end{pmatrix}
\end{equation}
where the coefficients are given by $P = -i[(J_{x} + J_{y})t_{\textbf{k}} - 2\mathfrak{z}(1 + J_{2})J_{z}]$, $Q = -i(J_{x} - J_{y})t_{\textbf{k}}$,  $\mathfrak{z}=4$ is the coordinate number, $t_{\textbf{k}} = 2\cos(k_{x}a) + 2\cos(k_{y}a) + J_{2}[e^{i(k_{x} + k_{y})a} + e^{i(k_{x} - k_{y})a} + e^{-i(k_{x} - k_{y})a} + e^{-i(k_{x} + k_{y})a}]$, and $a$ is the lattice constant \cite{TonyELee2011PRA}.

The steady state is dynamically stable when the real parts of all the eigenvalues of $\mathcal{L}_{\textbf{k}}$ are negative, otherwise it is unstable to the perturbation of wave vector $\textbf{k} = (k_{x},k_{y})$.
We define the most unstable eigenvalue $\lambda_{\text{max}}$ as the one with largest positive real part; the wave vector $\textbf{k} = (k_{x},k_{y})$ associated to the most unstable eigenvalue can be used to distinguish distinct phases \cite{XingLi2021PRB,MCCross1993RMP,AlexandreLeBoite2013PRL,AlexandreLeBoite2014PRA,JiasenJin2016PRX}.

Additionally, we choose the initial states for each sublattice as $|\psi_{A}(0)\rangle = (|\uparrow\rangle+|\downarrow\rangle)/\sqrt{2}$, $|\psi_{B}(0)\rangle = (|\uparrow\rangle+e^{i\pi/2}|\downarrow\rangle)/\sqrt{2}$, $|\psi_{C}(0)\rangle = (|\uparrow\rangle+e^{i\pi}|\downarrow\rangle)/\sqrt{2}$ and $|\psi_{D}(0)\rangle = (|\uparrow\rangle+e^{i3\pi/2}|\downarrow\rangle)/\sqrt{2}$ to investigate the time-evolution of the system, although the steady states are independent of the initial states.

\subsection{FM phase}
We start with the phase transition from PM to FM phase. The critical point for the PM-FM phase transition can be obtained by solving Eqs. (\ref{Mean-field equation system}), the explicit expression is given by
\begin{equation}
J_{x, y}^{c}=\frac{1}{16\mathfrak{z}^2(1 + J_2)^2(J_z-J_{y,x})}+J_z.
\label{cp_PMFM}
\end{equation}

\begin{figure}
\centering
\begin{minipage}[b]{1\linewidth}
\includegraphics[width=1\linewidth]{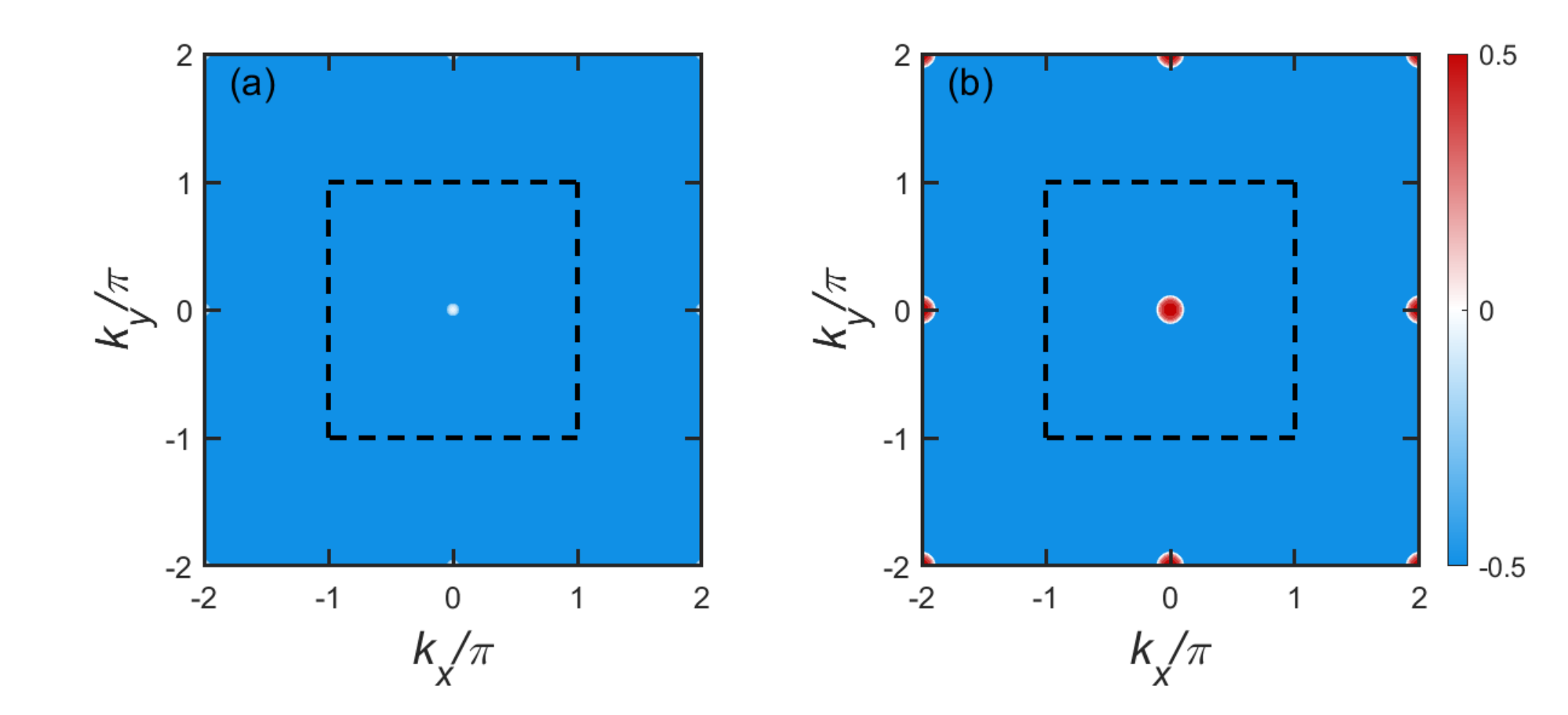} \\
\includegraphics[width=1\linewidth]{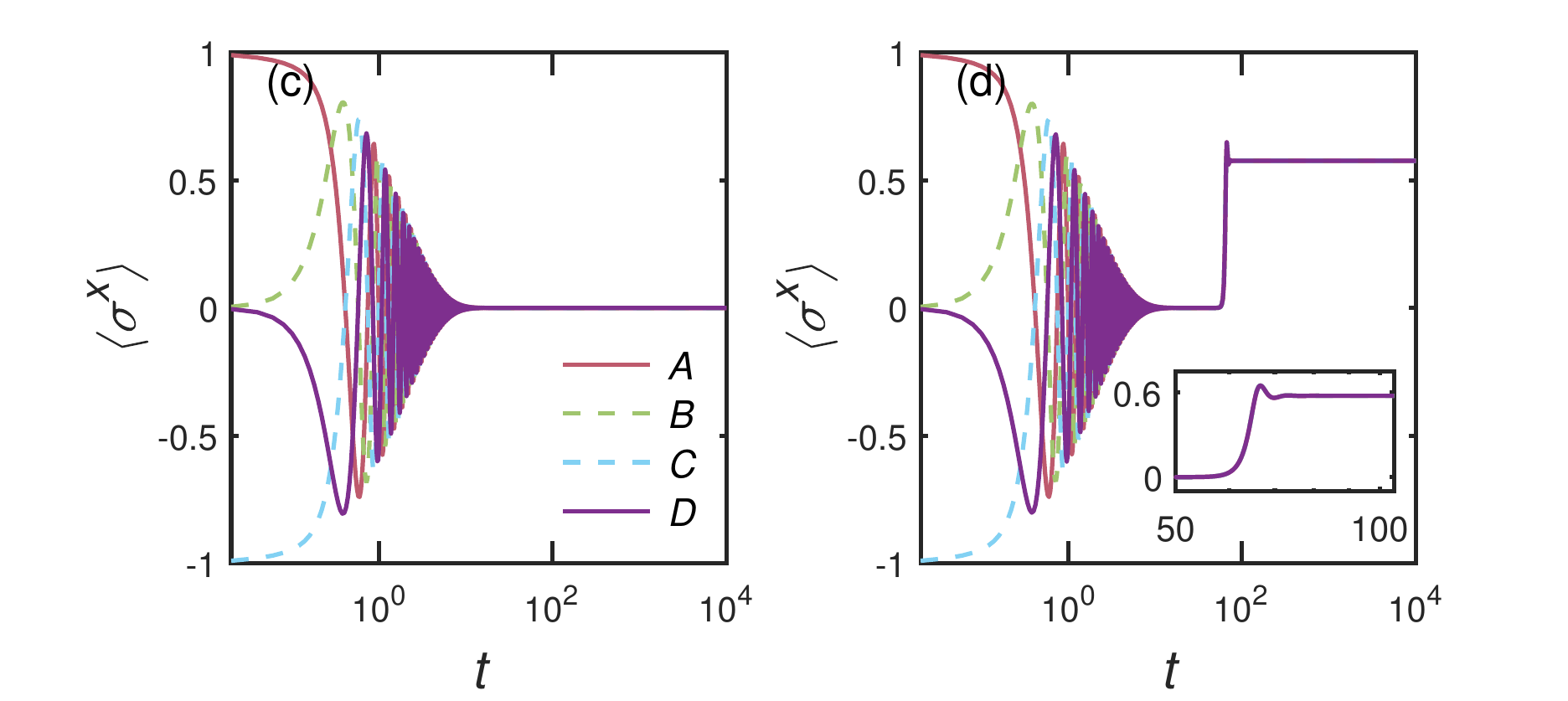}
\caption{\label{fig2} (Color online) (a) and (b) The real part of the most unstable eigenvalue as a function of the wave vectors $k_{x}$ and $k_{y}$ in momentum space. The dashed line indicates the first Brillouin zone. (c) and (d) The MF steady-state magnetizations along $x$-direction for the sublattices.
The parameters are chosen as $J_2=0.9$, $J_y = 0.9$, $J_z=1$ and $J_x=1.01$ for panels (a) and (c), 1.05 for panels (b) and (d). The labels of sublattices are consistent with Fig. \ref{illustration}.}
\end{minipage}
\end{figure}

In Fig. \ref{fig2}, we show the time-evolution of the magnetization $\langle\sigma^x(t)\rangle$ for $J_x=1.01$ (PM phase) and $1.05$ (FM phase) and $J_y=0.9$. In the PM phase, the magnetizations of all the sites approach to zero after sufficiently long time, regardless of the initial magnetization. While in the FM phase, after a transient oscillation the state of each site firstly evolves a metastable region with almost vanishing $\langle\sigma^x(t)\rangle$ and eventually ends up in the steady state with nonzero magnetization along $x$-direction. The appearance of the metastable state is because the chosen coupling parameter is close to the critical point.

In order to give the intuitive pictures for the PM and FM phases, we show the real parts of the most unstable eigenvalues  in the momentum space in Fig. \ref{fig2}. For $J_x=1.01$, the real part of $\lambda_{\text{max}}$  is always negative in the $k_x$-$k_y$ plane indicating that $\hat{\rho}_{\downarrow}$ is stable against the perturbations. The maximum of the real part is about $-0.0193$ locating at the origin of the momentum space. As the coupling strength $J_x$ increasing, the maximum of the real part increases and become positive in the FM phase. It is shown in Fig. \ref{fig2} that the maximum is positive for $J_x=1.05$ and the position of the maximum remains at the origin. In this case, the state $\hat{\rho}_{\downarrow}$ is unstable against uniform perturbations which offset each spin with a nonzero magnetization on the $x$-$y$ plane. This indicates the appearance of FM phase.

\subsection{The PM-CAF transition}
In this subsection, we discuss the phase transition from PM to antiferromagnetic phases. Similarly to Eq. (\ref{cp_PMFM}), one can obtain the expression for the critical point for PM-CAF transition as
\begin{equation}
J_{x, y}^{c}=-\frac{1}{16\mathfrak{z}J_2[(1 + J_2)J_z+J_2J_{y,x}]} - \frac{1+J_2}{J_2}J_z.
\end{equation}

\begin{figure}
\centering
\begin{minipage}[b]{1\linewidth}
\includegraphics[width=1\linewidth]{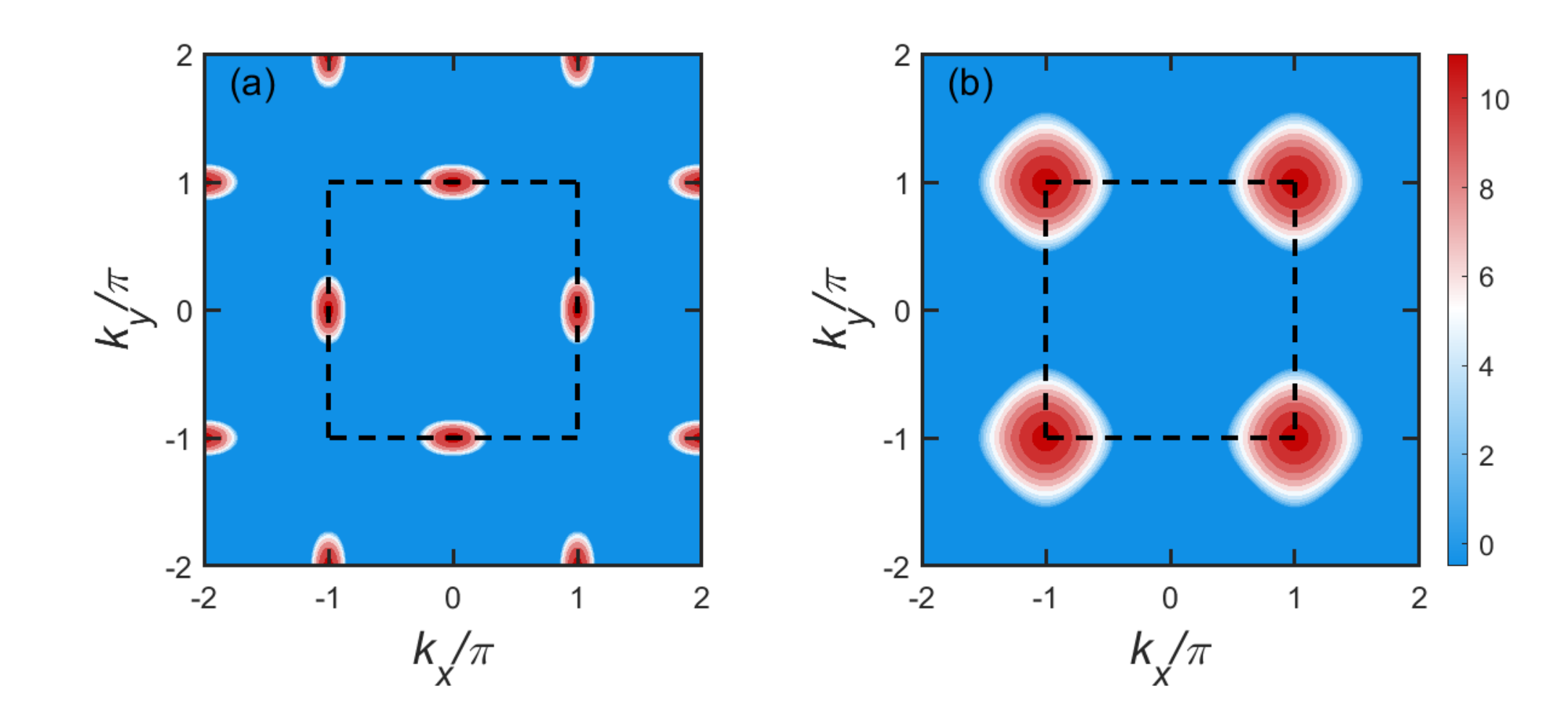} \\
\includegraphics[width=1\linewidth]{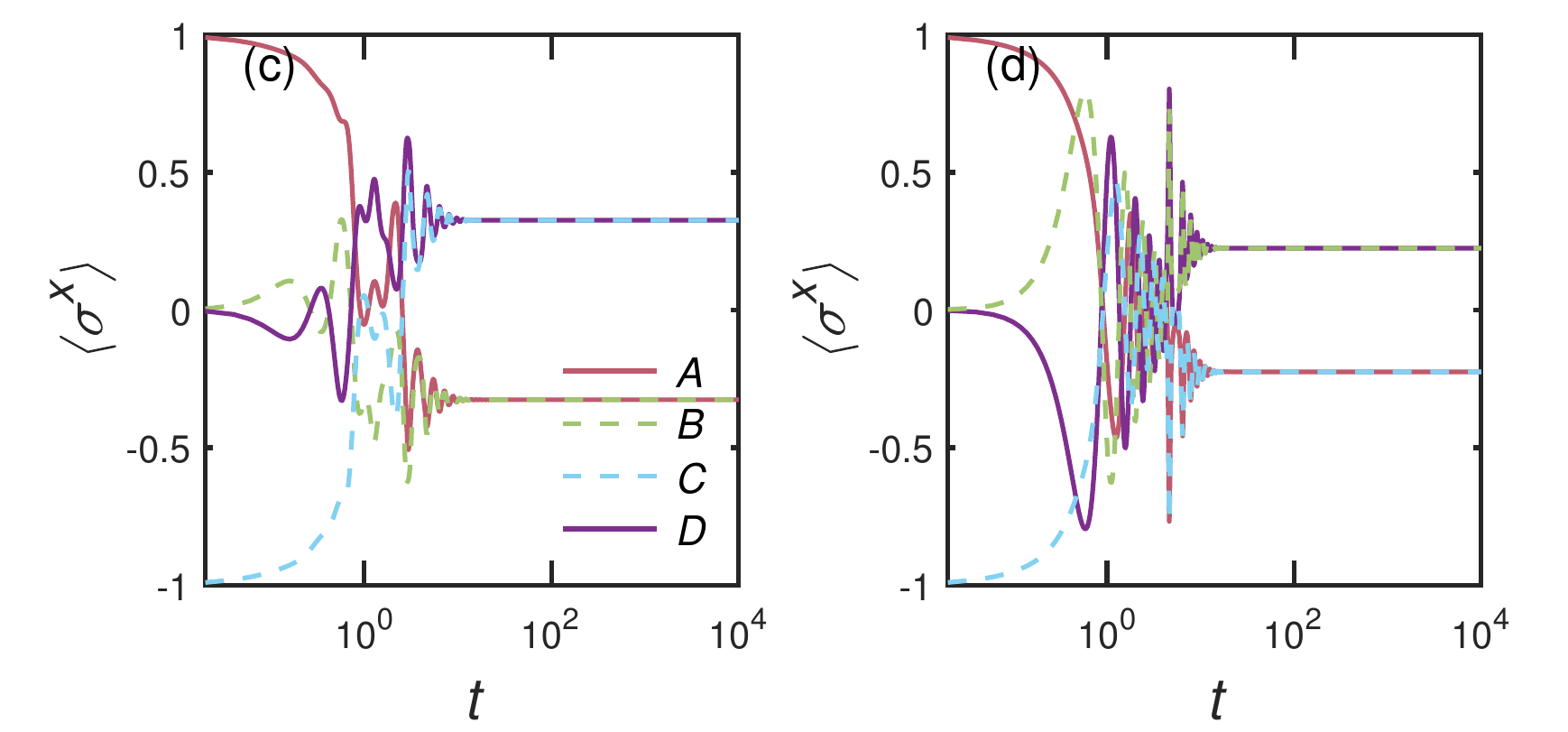}
\caption{\label{fig3}(Color online) (a) and (b)The real part of the most unstable eigenvalue in momentum space. The dashed line indicates the first Brillouin zone. (c) and (d) The MF steady-state magnetizations along $x$-direction for the sublattices. The parameters are chosen as $J_{x} = -2.5$, $J_{y} = 0.9$, $J_{z} = 1$ and $J_{2} = 0.9$ for panels (a) and (c) and $J_{2} = 0.1$ for panels (b) and (d). The labels of the sublattices are consistent with Fig. \ref{illustration}.}
\end{minipage}
\end{figure}

The real part of the most unstable eigenvalue in the momentum space in the CAF phase is shown in Fig. \ref{fig3}(a). The appearance of positive-valued maximum at $\textbf{k}=(0,\pm\pi)$ and $(\pm\pi,0)$ indicates that $\hat{\rho}_{\downarrow}$ is unstable against perturbations in terms of plane wave along $x$ or $y$ direction. Such perturbations give rise to a spatial modulation of the magnetizations along $x$ or $y$ direction with wavelength being twice the lattice constant. In Fig. \ref{fig3}(c), the time-evolution the magnetization $\langle\hat{\sigma}^x(t)\rangle$ for each sublattice are shown. One can see that the magnetizations $\langle\hat{\sigma}^x\rangle$ of the sublattices in the same column evolve to the same steady-state value for sufficient long time indicating the CAF pattern.

Interestingly, we find that the CAF phase may become AFM phase by varying the strength of NNN coupling $J_2$. The real part of $\lambda_{\text{max}}$ in the momentum space in the AFM phase with $J_2=0.1$ is shown in Fig. \ref{fig3}(b). Compared to the case of CAF phase, in the AFM phase the positive maximum appears at $\textbf{k}=(\pm\pi,\pm\pi)$ (the high-symmetry point $M$ in the first Brillouin zone). This corresponds to the perturbation in both $x$ and $y$ directions. The steady-state magnetization is modulated in both directions with a period of two lattice sites; the whole lattice is actually divided into two sublattices. The steady-state pattern of the AFM state is also revealed by the time-evolution of the magnetization $\langle\sigma^x\rangle$. From Fig. \ref{fig3}(d) one can see that magnetizations in the long-time limit exhibit $\langle\sigma^x_A\rangle=\langle\sigma^x_C\rangle\ne\langle\sigma^x_B\rangle=\langle\sigma^x_D\rangle$.

We recall that for $J_2=0$, the model reduces to the conventional XYZ model in which only AFM phase exists \cite{TonyELee2013PRL}. In this sense, the CAF phase can be considered as a result of the presence of NNN interaction and the competition to the NN interaction.

\section{Cluster Mean-field Method}
\label{Sec:Cluster Mean-field Method}
So far, we have neglected all the correlations in the discussion. In order to refine the MF results, we will take the short-range correlation into account in the analysis. To this aim, we apply the CMF technique to our model. In the CMF approximation,  as schematically shown in Fig.\ref{fig1}(e), the whole lattice is divided into a series of clusters $\mathcal{C}$ which is consisted of a number of contiguous sites. All the clusters are assumed to be identical. The density matrix of the whole lattice is thus factorized as the product of the density matrix of each cluster,
\begin{equation}
\hat{\rho}_{\text{CMF}} = \underset{\mathcal{C}}\bigotimes \hat{\rho}_{\mathcal{C}}.
\label{CMF factorization1}
\end{equation}
Substituting Eq. (\ref{CMF factorization1}) into Eq. (\ref{Lindblad Master Equation}) and taking the partial trace of the global density matrix over all the clusters except for $\mathcal{C}$, one can obtain the CMF master equation regarding to cluster $\mathcal{C}$ as the following
\begin{equation}
\frac{d\hat{\rho}_{\mathcal{C}}}{dt}=-i[\hat{H}_{\text{CMF}},\hat{\rho}_{\mathcal{C}}]+ \sum_{j\in\mathcal{C}}\mathcal{D}_{j}[\hat{\rho}_{\mathcal{C}}].
\label{CMF Masterequation}
\end{equation}
In the expression above, the CMF Hamiltonian is given by
\begin{equation}
\hat{H}_{\text{CMF}} = \hat{H}_{C} + \hat{H}_{\mathcal{B}(\mathcal{C})},
\label{Hamiltonian_CMF}
\end{equation}
where $\hat{H}_C=\sum_\alpha{J_\alpha\left[J_1\sum_{\langle j,l\rangle\in C}{\hat{\sigma}_j^\alpha\hat{\sigma}_l^\alpha}+J_2\sum_{\langle\langle j,l\rangle\rangle\in C}{\hat{\sigma}_j^\alpha\hat{\sigma}_l^\alpha}\right]}$, ($\alpha=x,y,z$, $\langle\cdot,\cdot\rangle$ and $\langle\langle\cdot,\cdot\rangle\rangle$ denote the NN and NNN sites) describes interactions between the sites inside the cluster $\mathcal{C}$, while $\hat{H}_{\mathcal{B}(\mathcal{C})}=\sum_{\alpha}{J_\alpha\left[J_1\sum_{\langle j,l\rangle}{\hat{\sigma}^\alpha_j\langle\hat{\sigma}^\alpha_l\rangle}+J_2\sum_{\langle\langle j,l\rangle\rangle}{\hat{\sigma}^\alpha_j\langle\hat{\sigma}^\alpha_l\rangle}\right]}$ where $j\in\mathcal{C}$ and $l\in\mathcal{C'}$ ($C'$ is the cluster adjacent to $\mathcal{C}$) describes the inter-cluster interactions. More details about the CMF approximation can be found in Ref. \cite{JiasenJin2016PRX}.

As shown in Eq. (16), the idea of CMF approximation
is that the interactions between the sites inside a cluster are
treated exactly, while the interactions between different clusters
are treated at the mean-field level. In principle, as the size of cluster approaching to infinity the correlations embedded in the lattice are gradually in included in the analysis, we are able to obtain the property of the system in thermodynamic limit. The conventional MF approximation is considered to be a limit case for which all the correlations are neglected.

Here, we will use a series of rectangular clusters of size $L=n_1\times n_2$ in the CMF analysis. To be specific, for the clusters of size $L\le9$ we employ the standard fourth-order Runge-Kutta method to directly integrate the CMF master equation Eq.(\ref{CMF Masterequation});  for $L>10$ we combine the CMF approximation with the quantum trajectory method \cite{moler1993,JeanDalibard1992PRA,MBPlenio1998RMP} and the results are obtained by averaging over 500 trajectories \cite{JiasenJin2016PRX}.

We would like to note that such rectangular clusters are convenient in revealing the FM or AFM nature of the steady states in our model since their translation invariance along both $x$ and $y$ directions. Moreover, this choice simplifies the determination of the effective field in $\hat{H}_{\mathcal{B}(\mathcal{C})}$. Series of clusters in other forms may also be used for CMF analysis, for instance, the clusters that can tile the square lattice \cite{dagotto1994}. The steady-state property of the system in thermodynamic limit is independent of the choice of clusters.

\begin{figure}[!htp]
\includegraphics[width=0.49\textwidth]{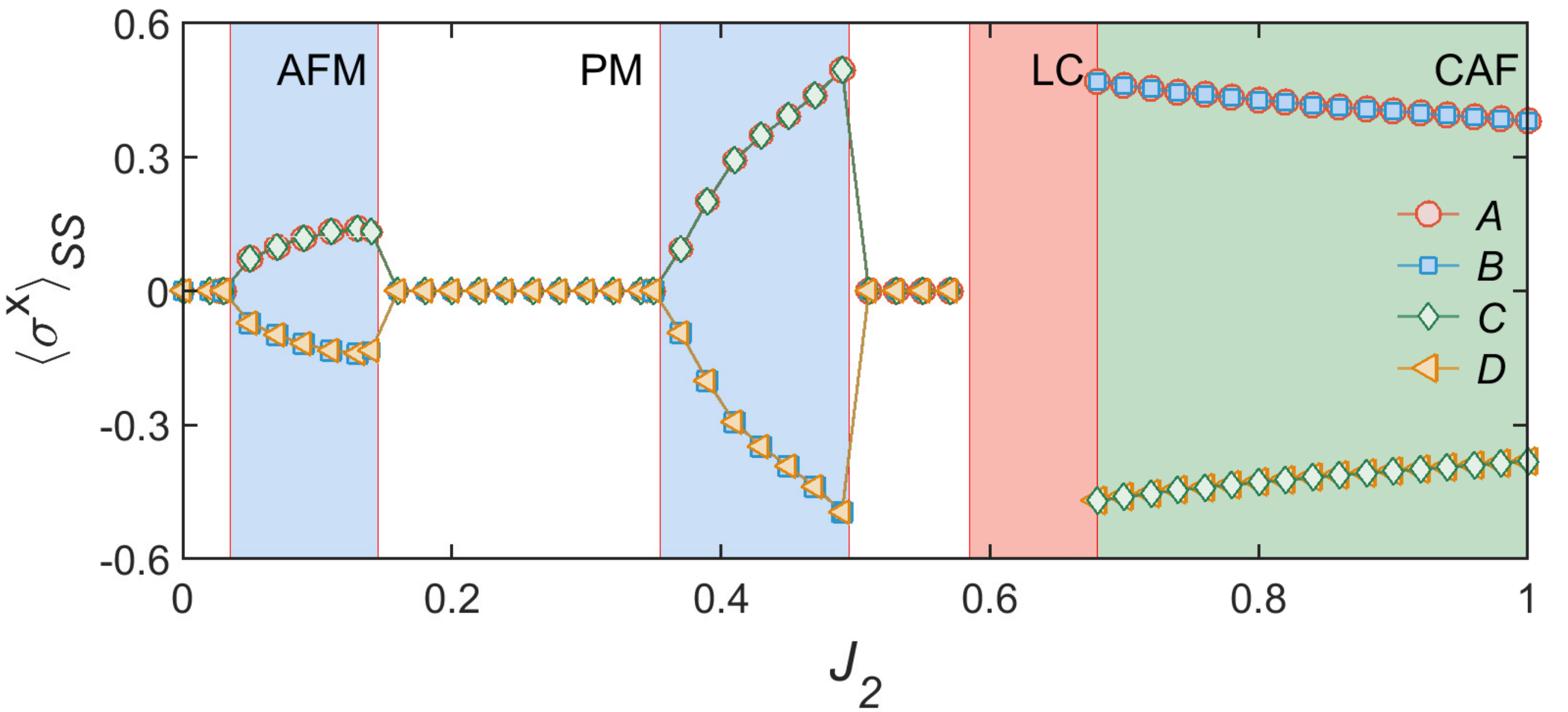}
\caption{\label{fig4} (Color online) The steady-state magnetizations along $x$-direction for a $2\times2$ cluster (each site indicated by various symbol) as a function of NNN coupling strength $J_2$. In the limit cycle region, the data is lacking since the system never reach the asymptotical steady state in long-time limit. With $J_2$ increasing, the system exhibits the \text{PM}, \text{AFM}, \text{LC}, \text{CAF} steady-state phases.}
\end{figure}

\subsection{CAF phase}
In Fig.\ref{fig4}, we show the CMF phase diagram as a function of $J_2$ with a cluster of size $L = 2\times2$, the other parameters are chosen as $\{J_x,J_y,J_z\} = \{-3.2,-1,1\}$. One can see that as the strength of NNN coupling increasing the system exhibits various steady-state phases. For the limit cases of small and large $J_2$, the steady-state phases are similar to those in the equilibrium system.

To corroborate the existence of the CAF phase, we investigate the steady-state magnetizations by systematically increasing the size of the clusters. In accordance with the definition of CAF phase, we choose the order parameter as $\mathcal{O}_{\text{CAF}} = \sum^{L}_{j=1}\sum_{\langle\langle j,l\rangle\rangle}|\langle\sigma^{x}_{j}\rangle_{ss}-\langle\sigma^{x}_{l}\rangle_{ss}|/\ell$, with $\ell$ is the total number of NNN interactions. This order parameter shows the steady-state magnetization difference between the $j$-th site and its NNN site. The nonzero value of $\mathcal{O}_{\text{CAF}}$ indicates the system is in the CAF phase.

In Fig. \ref{fig5} we show the order parameter $\mathcal{O}_{\text{CAF}}$ as a function of $J_{2}$ in the CMF approximation with different sizes of clusters. For the chosen anisotropic coupling case, both the single-site MF and CMF results show the existence of the CAF phases at larger $J_2$. As the size of cluster increasing, the CAF phase shrinks and remains in $0.7\lesssim J_2 < 1$ up to $4\times4$ cluster. One point should be clarified is that the order parameter $\mathcal{O}_{\text{CAF}}$ in the regions other than CAF phase is not exactly zero for the $2\times4$ cluster is because of the geometrical anisotropy of the cluster.

\begin{figure}[!htp]
\includegraphics[width=0.45\textwidth]{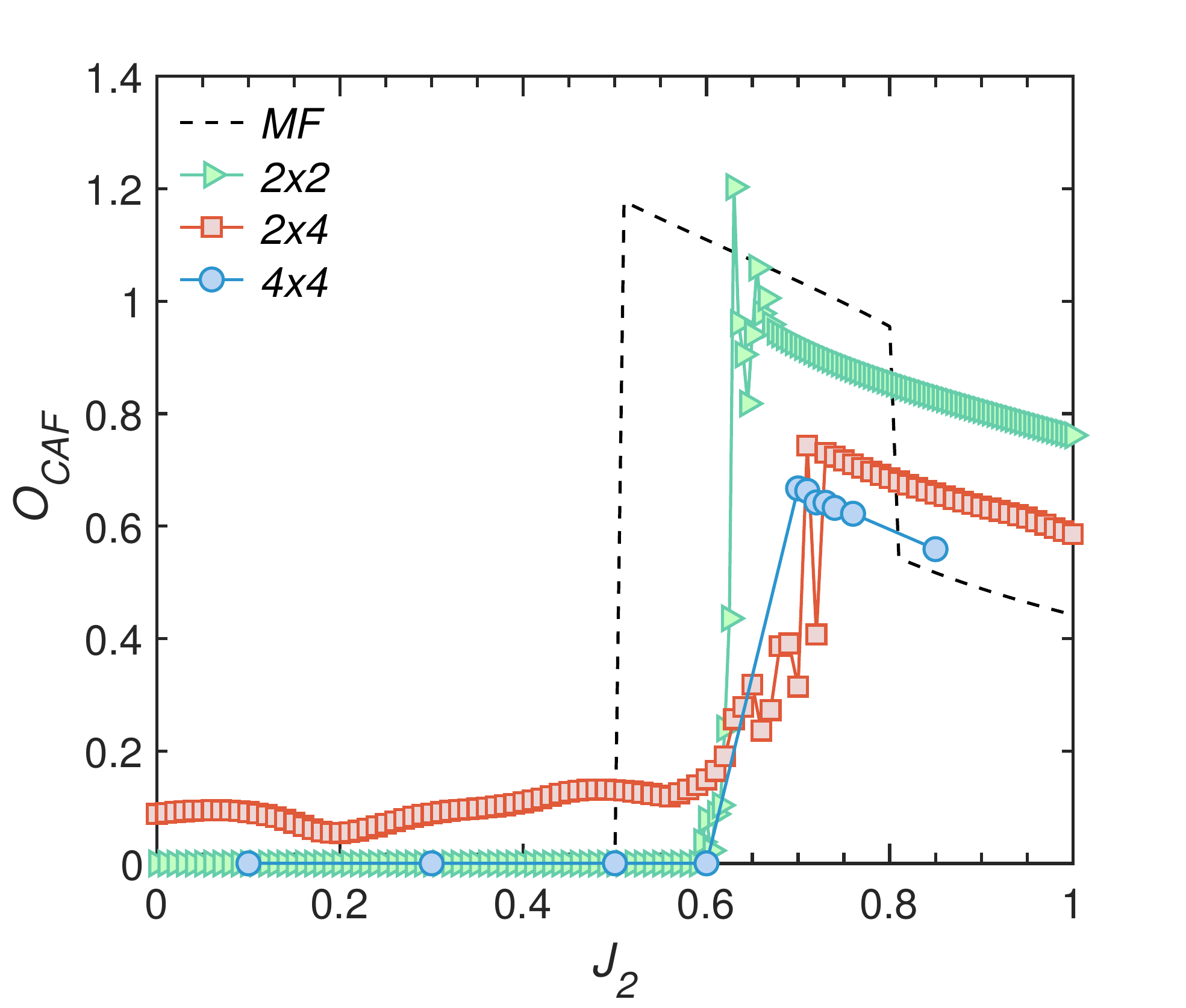}
\caption{\label{fig5}(Color online) The order parameter $\mathcal{O}_{\text{CAF}}$ as a function of $J_{2}$ for various sizes of clusters. Other parameters are chosen as $\{J_x,J_y,J_z\} = \{-3.2,-1,1\}$. The instable CMF data in $0.6\le J_{2}\le 0.7$ means the steady state of the system is limit cycles. .}
\end{figure}

\subsection{Limit Cycle}
Interestingly, in the $2\times2$ CMF approximation a limit cycle region emerges for $0.59\lesssim J_2\lesssim0.63$. In the limit cycle region, the magnetization of each site oscillates periodically with time instead of reaching an asymptotic steady state in the long-time limit. The limit cycles are common in classical nonlinear dynamical systems and feature a stable closed trajectory in phase space. For the dissipative spin-1/2 XYZ model with only NN interaction, the limit cycle is predicted by the single-site MF approximation \cite{chan2015,ETOwen2018NJP}. On the contrary, here in our model with frustrated interactions, although the limit cycle is missed by the single-site MF approximation, it is uncovered by the inclusion of short-range correlations in the CMF approximation.

In order to discriminate that wether the time-dependent oscillation of the magnetization is a limit cycle or chaos, we employ the so-called largest Lyapunov exponent as proposed in Ref.\cite{Yusipov2019chaos}.
\begin{figure}[!htp]
\includegraphics[width=0.51\textwidth]{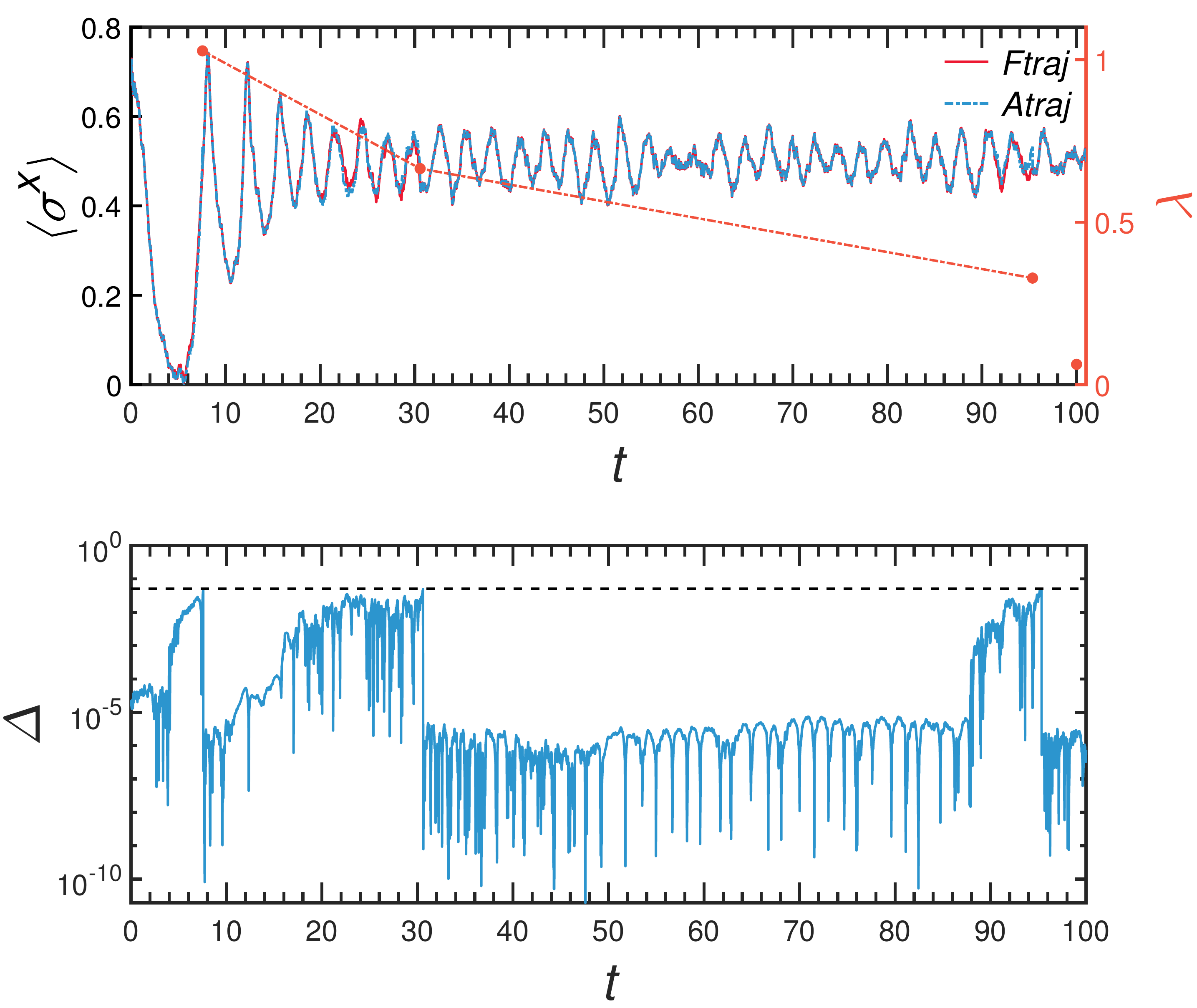}
\caption{\label{fig6}(Color online) The top panel shows the time-dependent magnetization $\langle \sigma^{x}\rangle$ of the fiducial trajectories (Ftraj) and auxiliary trajectories (Atraj). The red discrete dots represent the changing of the largest
LE over time. The bottom panel shows changes of the difference of the observable $\Delta$ as a function time.  The black dash line denotes the threshold $\Delta_{\text{max}}=0.05$ for resetting the auxiliary trajectory.}
\end{figure}

Analogous to the classical definition, we use the quantum trajectories to simulate the evolution of the system. The largest LE is thus defined by the ``distance" between the fiducial and auxiliary trajectories. The ``distance" may be obtained by direct calculation of the difference of  some observables. In this work, we choose the observable to be the magnetization along $x$ direction $\bar{\sigma}^{x}(t)$ , which can be obtained by
\begin{equation}
\bar{\sigma}^{x}(t) = \frac{1}{N}\sum^{N}_{j=1}\langle\psi_{j}(t)|\hat{\sigma}^{x}|\psi_{j}(t)\rangle.
\end{equation}

The fiducial trajectory is initialized as a normalized quantum state vector $\psi^{\text{ini}}_{f}$, and the auxiliary one is also prepared as a normalized state with a perturbation on the fiducial initial state,
\begin{equation}
\psi^{\text{ini}}_{a}=\frac{\psi^{\text{ini}}_{f} + \delta\psi_{p}}{||\psi^{\text{ini}}_{f} + \delta\psi_{p}||}.
\end{equation}
Here $\psi_{p}$ is a random perturbative state and $\delta\ll 1$. The difference of obseravables $\Delta(t) = |\bar{\sigma}^{x}_{f}(t)-\bar{\sigma}^{x}_{a}(t)|$ is time-dependent. The initial value can be calculated by $\Delta_{0}=|\bar{\sigma}^{x}_{f}(0)-\bar{\sigma}^{x}_{a}(0)|$, If the difference exceeds threshold $\Delta(t_{k})>\Delta_{\text{max}}$ at the time $t_{k}$, the growth factor of the largest LE $d_{k}=\Delta(t_{k})/\Delta_{0}$ is summed and the auxiliary state vectors have to be renormalized close to the fiducial trajectories. The difference of observables is reset to the initial value. The largest LE can be estimated as
\begin{equation}
\lambda = \lim_{t\to\infty}\frac{1}{t}\sum_{k}\ln d_{k}.
\end{equation}

Here we discuss the cluster of size $L=2\times2$. In the simulation, the number of trajectories is $M =300$ and the threshold is $\Delta_{\text{max}}= 0.05$. The numerical result is shown in Fig. \ref{fig6}.  In the top panel, the left $y$-axis corresponds to the time-evolution of on-site magnetization for fiducial and auxiliary trajectories, $\bar{\sigma}^{x}_{f}(t)$ and $\bar{\sigma}^{x}_{a}(t)$, respectively. Although the amplitude of the oscillation fluctuates due the probabilistic nature of the quantum trajectory method, the significant oscillating behavior can be observed. The right $y$-axis is related to the change of the largest LE with evolution time. The largest LE is updated three times in the time interval of $t\in[0,100]$, corresponding to the time at which the difference exceeds the threshold. We have extended the simulation to $t = 500$,  the largest LE descends from $\{\lambda,t\}=\{1.0269,7.5645\}$ to $\{\lambda,t\}=\{0.0625,500\}$. The largest LE at $t=500$ is shown by the most-right orange symbol.

The bottom panel is the time-evolution of the difference of observables $\Delta(t)$. It can be seen that there are three discontinuous jumps at $t\approx8$, $30$, and $96$. The jumps mark the events that the difference of observables exceeds the threshold. The difference of at $t=500$ is $\Delta<3\times10^{-10}$ (not shown) which is small enough to indicate that the largest LE will continue descend in the long-time limit. We can conclude that the $\Delta(t)$ will eventually reach zero and the oscillation is stable.

\begin{figure}[!htp]
\includegraphics[width=0.5\textwidth]{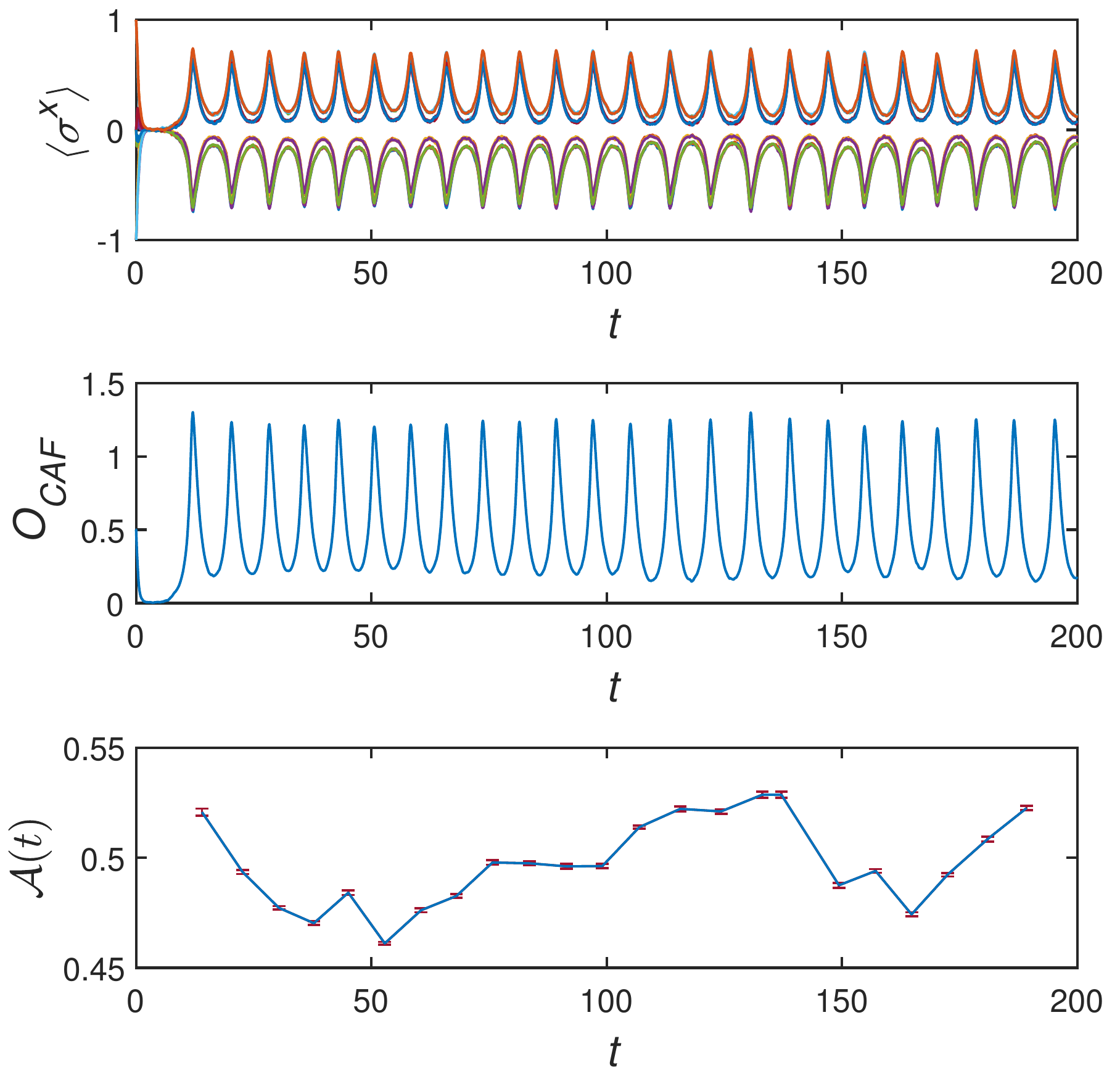}
\caption{(Color online) The top panel presents the time-dependent magnetization $\langle \sigma^{x}_{j}\rangle$ for a $4\times4$ cluster. The middle panel shows the time-dependent order parameter $\mathcal{O}_{\text{CAF}}$, and the time-dependent average amplitude is shown in the bottom panel, the error bars denote the variances of the average amplitude, and the parameters are chosen as $J_x = -3.2,J_y = -1,J_z = 1$, and $J_2 = 0.63$.}
\label{fig7}
\end{figure}

In the upper panel of Fig.\ref{fig7}, we show the time-evolution of the magnetization for each sublattice in the limit cycle region with the $4\times4$ CMF approximation. The time-dependence of the CAF order parameter is shown in the middle panel as well. One can see that the magnetization is time-dependent and showing a CAF pattern. As reported in Refs. \cite{chan2015,ETOwen2018NJP}, the limit cycle is absent in the case of $J_2=0$, the coexistence of the CAF ordering and oscillating magnetization is combination of the effects of the interaction frustration and nonequilibrium nature of the system. To check the stability of the oscillation, We define $\mathcal{A}(t)=\sum^{L}_{j=1}\mathcal{A}[\langle\hat{\sigma}_j^{x}\rangle]/L$, where $\mathcal{A}[\langle\hat{\sigma}_j^{x}\rangle]$ measures the difference between the local maximum and minimum values. The averaged amplitude $\mathcal{A}(t)$ which is shown in the bottom panel of Fig.\ref{fig7}. The error bars are the variances of the average oscillation amplitude. In each oscillation period, the peaks or the valleys of the time-dependent magnetization of each site do not always accurately locate at the same time. After determining the local maximum $t^{\text{loc}}_{\text{max}}$ or minimum time point $t^{\text{loc}}_{\text{min}}$ for each magnetization, we average the each magnetization over a small time window $\Delta t$ to obtain the amplitude, e.g. $\Delta t = t^{\text{loc}}_{\text{max}}\pm\delta t, \delta t = 0.4$. Without considering the fluctuations $\mathcal{A}(t)$ caused by the probabilistic nature of the quantum trajectory method, the value of $\mathcal{A}(t)$ stays in the range $0.45\le \mathcal{A}(t) \le 0.55$ indicating that the oscillation is stable.

\section{Summary}
\label{Sec:Conclusions}
In summary, we have investigated the steady-state phases of the dissipative spin-1/2 XYZ model with $J_1$-$J_2$ couplings. Compared with the previous studies on the same model but with only NN couplings ($J_1$), the presence of the interaction frustration induced by the NNN coupling ($J_2$) indeed enriches the steady-state phases. In order to study the dynamics of the system, we perform the mean-field approximation, basing on the Gutzwiller factorization, to decouple the master equation that governs the dynamics of the whole lattice. We check the linear stability of the fixed points to the system of single-site MF Bloch equations. The results from the single-site mean-field approximation reveal the emergence of the various antiferromagnetic phases, including the AFM and CAF phases. The critical point of the PM-CAF phase transition is presented.

The formalism of the linear stability analysis reminds us of the well-known spin-wave theory in determining the low-energy excitation of the magnetically ordered system. In the spin-wave theory, the spins of the considered system are assumed to be aligned in the same direction. The wave-like low-energy excitation is created by the spin operator in the reciprocal lattice $\hat{S}_{\textbf{k}}=\sum_{j}{e^{i\textbf{k}\cdot\textbf{r}_j}\hat{S}_j}$. The thermodynamics properties as well as the dynamics of the considered system can thus be investigated by the diagonalization of the Hamiltonian in the momentum space.
Recently, the spin-wave approximation has been used in studying the effect of the external magnetic field in quantum spin system which is described by the Karplus-Schwinger master equation \cite{zvyagin2020}. The application of spin-wave approximation to open quantum systems would be an interesting topic.

The existence of the antiferromagnetic phases in the thermodynamic limit is confirmed by a series of CMF analysis. The short-range correlations are gradually included as the size of clusters increasing. The CAF order remains nonzero up to the $4\times4$ CMF approximation. Moreover, we find the evidence of the LC phase, in which the system is in a time-periodic oscillating state in the long-time limit, in the CMF approximation. The fact that the LC phase is absent in sing-site MF approximation but appears in the CMF approximation is the unique feature in the interaction frustrated system. The investigations on the largest quantum Lyapunov exponent and the averaged oscillation amplitude support the existence of the LC phase up to $4\times4$ CMF approximation.

Finally, we note that the properties of the steady-state
phases investigated in this paper, especially the stability of the
LC phase, are limited by the cluster size; analysis on the larger
size cluster is still required. On this perspective, the combination of the CMF approach with other available techniques are promising to achieve this purpose, such as machine learning techniques \cite{MichaelJHartmann2019PRL,AlexandraNagy2019PRL,FilippoVicentini2019PRL,NobuyukiYoshioka2019PRB,yuan2021} and the corner-space renormalization \cite{SFinazzi2015PRL,RiccardoRota2019PRL}. A comprehensive panorama of the simulation methods for open quantum many-body systems can be found in Ref. \cite{weimer2021}. In addition, our theoretical predictions may be experimentally investigated in following different platforms\cite{DPorras2004PRL,RMelzi2001PRB,Hajime2007PRB}.

\section*{ACKNOWLEDGMENTS}
This work is supported by National Natural Science Foundation of China via Grant No. 11975064.

\end{document}